\documentclass[prb,twocolumn,showpacs]{revtex4}% Physical Review B

\usepackage{graphicx}
\usepackage{dcolumn}
\usepackage{amsmath}

\begin{document}

\preprint{HEP/123-qed}

\title{Linewidth of single photon transitions in Mn$_{12}$-acetate}

\author{Beth Parks}
\email{meparks@mail.colgate.edu}
\author{Joseph Loomis}
\affiliation{Department of Physics and Astronomy, Colgate University \\
Hamilton, NY 13346}
\author{Evan Rumberger}
\author{David N. Hendrickson}
\affiliation{Department of Chemistry and Biochemistry, University of 
California, San Diego \\
La Jolla, CA 92093-0358}
\author{George Christou}
\affiliation{Department of Chemistry, Indiana University \\
Bloomington, IN 47405}

\date{\today} 

\begin{abstract} 
We use time-domain terahertz spectroscopy to measure the position and 
linewidth of single photon transitions in Mn$_{12}$-acetate.  This 
linewidth is compared to the linewidth measured in tunneling experiments.  
We conclude that local magnetic fields (due to dipole or hyperfine 
interactions) cannot be responsible for the observed linewidth, and 
suggest that the linewidth is due to variations in the anisotropy 
constants for different clusters.  We also calculate a lower limit on 
the dipole field distribution that would be expected due to random 
orientations of clusters and find that collective effects must narrow 
this distribution in tunneling measurements.
\end{abstract}

\pacs{76.30.-v,75.45.+j,78.30.-j}

\maketitle

Mn$_{12}$-acetate ([Mn$_{12}$O$_{12}$(CH$_3$COO)$_{16}$(H$_2$O)$_4$]
$\cdot$ 2~CH$_3$COOH$\cdot$4H$_2$O) 
is a member of a class of high-spin molecular clusters that have
been shown to exhibit quantum tunneling of the magnetic
moment.\cite{Friedman96,Thomas96}  It consists of a core of twelve
manganese ions with spins tightly coupled via superexchange through
twelve oxide ions, with a ground-state spin $S = 10$.  These clusters are
separated by acetate and water groups and arranged in a tetragonal
body-centered lattice so that the nearest neighbor distance is
13.7~${\rm \AA}$ and the shortest distance between manganese ions in
neighboring clusters is 7~${\rm \AA}$.  Since it is fairly unusual for a
system this large to display quantum mechanical properties, it has been
the subject of much investigation.

One particularly interesting area of investigation is the interaction of
the spins with their environment, as reflected in the linewidth of the
energy levels.  It is not yet clear why in some cases the measured
linewidth of transitions provides information about the intrinsic
properties of the clusters (homogeneous broadening), while in others it
seems to be due to variations in the local environments of clusters
(heterogeneous broadening).  A comparison of the linewidths we measure
for intrawell transitions with the linewidths measured in tunneling
between wells can help to determine the answer.

The Hamiltonian for the spin clusters is approximately given by $\cal{H}
= $$-\alpha S_{z}^{2} - \beta  S_{z}^{4} + \gamma (S_{+}^{4} +
S_{-}^{4}) - g \mu_{B} \bf{S} \cdot \bf{H}$, where $\alpha$ = 0.38~
cm$^{-1}$, $\beta = 8.2 \times 10^{-4}$~cm$^{-1}$, $\gamma \sim \pm 2
\times 10^{-5}$~cm$^{-1}$, and $g \sim 2$.
\cite{Barra97,Mirebeau99,Hill98,Fort98,Luis98,Bao01,Mukhin,Zhong99} In
zero field, states with equal $|m|$ are degenerate.  The ground states
$m = \pm 10$ are separated by a barrier of approximately 66~K.

In this paper, we describe an experiment that measures the linewidth of
the intrawell $m=10 \rightarrow$~9 transition (and the -10~$\rightarrow$~
-9 transition) using time-domain terahertz spectroscopy.  The
measurements were made on a pellet pressed from small unaligned crystals
of Mn$_{12}$-acetate prepared according to the procedure of
T. Lis.\cite{Lis80}

In terahertz time-domain spectroscopy, a nearly single-cycle electromagnetic
pulse with a length of a few picoseconds is produced when an optical
pulse from a modelocked titanium sapphire laser is incident on a
microfabricated photoconductive antenna.  This electromagnetic pulse is
guided through the sample quasi-optically and then detected by a second
photoconductive antenna.  This detector antenna is gated by a
time-delayed laser pulse split from the same pulse that triggered the
generator antenna.  The time dependence of the transmitted
electromagnetic pulse is measured by adjusting the delay between the
laser pulses incident on the generator and detector.  The transmitted
electric field $E_{t}(t)$ is then Fourier transformed to yield
$\tilde{E}_{t}(\omega)$, the complex frequency dependence of the
transmitted electromagnetic pulse.  This transmitted electric field can
be normalized by the field measured with the sample removed from the
beam path, yielding the complex transmission coefficient,
$\tilde{t}(\omega)$.

Generally, terahertz spectroscopy is used to characterize a material
response that varies slowly with frequency, for
which the frequency resolution is not required to be higher than tens of
GHz.  To our knowledge, it has never been used to examine linewidths of
transitions.  This is because higher frequency resolution can only be
obtained with a longer time delay between the generator and detector
laser pulses.  It is difficult to align the translation stage well
enough to allow constant detection sensitivity over a large delay range,
since the detector is very sensitive to the position of the focussed
laser spot on the photoconductor.  Through a careful alignment process
we were able to maintain the detector sensitivity within a few percent
over the entire delay range.  It is also difficult to eliminate stray
reflections from surfaces of optical elements and ends of transmission
lines.  These stray reflections lead to interference fringes in the
spectrum.  The fringes make it impossible to fit the lineshape to a
particular form (Gaussian or Lorentzian), but it is still possible to
estimate the linewidth.

In Figure~1 we show the magnitude ratio of the transmission spectra for
the Mn$_{12}$-acetate pellet at 15~K and at 3~K.  This ratio makes the
changes in transmission due to temperature more apparent.  At 3~K,
almost all the spins are in the ground state ($m = \pm 10$), so only the
transitions from $m = \pm 10$ to $\pm 9$ are observed.  At 15~K, some
higher energy states are thermally occupied so more transitions are
observed.  These absorptions are seen as peaks in the transmission ratio.  
The location of these transitions can be fit to yield the
first two parameters in the Hamiltonian: $\alpha = 0.382(4)$~cm$^{-1}$ 
and $\beta = 8.0(2) * 10^{-4}$~cm$^{-1}$. These parameters are in good 
agreement with those cited earlier.

\begin{figure}
\includegraphics[width=\linewidth]{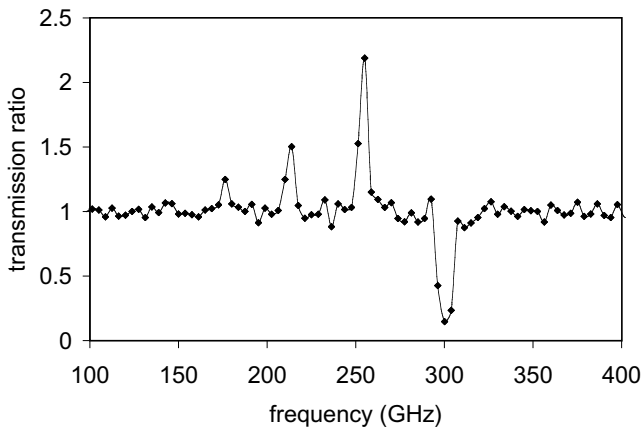}
\caption{Transmission magnitude at 3 K divided by the transmission 
magnitude at 15 K.  The peaks are due to transitions that are observed 
at 15 K but are not observed at 3 K because nearly all the clusters 
are in their ground states.}
\end{figure}

In Figure~2 we focus on the absorption at 300.6~GHz, which corresponds
to the transition from $m = \pm 10$ to $\pm 9$.  (The absorption from $m
= \pm 9$ to $\pm 8$ has a similar linewidth.)  We plot the index of
refraction as a function of frequency near this absorption at 
temperature T~=~2.1~K.  The index of refraction was calculated directly 
from the (complex) transmission spectrum using the following equation.  
The transmission through a slab of thickness $d$ = 1.4~mm with complex 
index of refraction $\tilde{n}$ 
is:\cite{Ibach} 
\begin{equation}
\tilde{t}(\omega) = \frac{4\tilde{n}}{(\tilde{n}+1)^2} 
\frac{e^{i \omega \tilde{n} d / c}}
{1-(\frac{\tilde{n}-1}{\tilde{n}+1})^{2} e^{2 i \omega \tilde{n} d / c}}.
\label{transmission}
\end{equation}

\begin{figure}
\includegraphics[width=\linewidth]{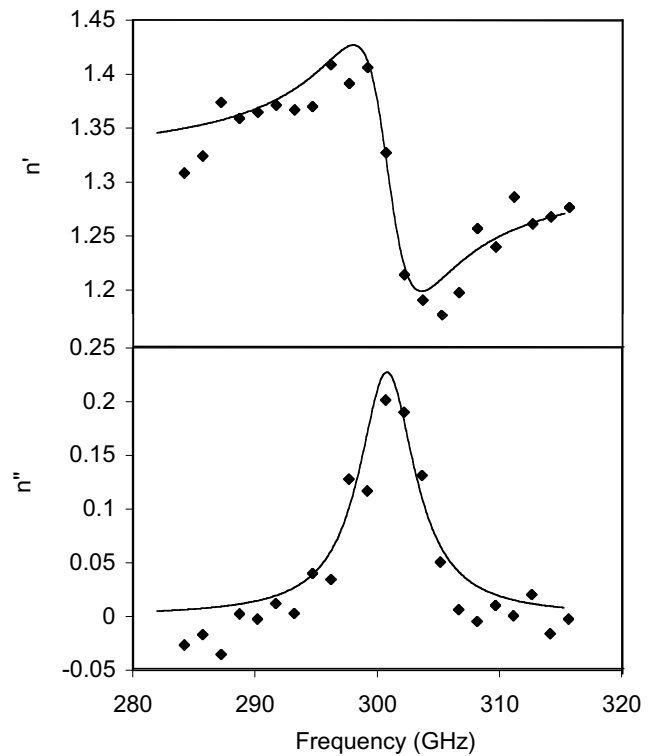}
\caption{Real and imaginary parts of the index of Mn$_{12}$-acetate.  
The index is calculated directly from the transmission using 
Eq.~(\ref{transmission}).  The line is a fit to Eq.~(\ref{resonance}).}
\end{figure}

This equation assumes that all reflections from the interfaces are
included, even though THz spectroscopy only collects data within a
finite time window after the main pulse.  However, in this case the
reflectivity is low enough and the absorption is strong enough that the
limited time window does not introduce significant error. Notice that
since this is a magnetic transition, $n$ must be correctly defined as
$\sqrt{\mu \epsilon / \mu_{0} \epsilon_{0}}$, rather than approximated
as $\sqrt{\epsilon / \epsilon_{0}}$.  We stress that our measurement
yields the complex index of refraction without any modeling of either
the lineshape or the high and low frequency extrapolations of the
response functions.

We model the absorption using the standard form for a magnetic dipole 
resonance:
\begin{equation}
\mu / \mu_{0} = 1 + \frac{F \omega_{0}^2}
{\omega_{0}^{2} - \omega^{2} - i \omega \Gamma}.
\label{resonance}
\end{equation}
We assume that $\epsilon$ is constant over this frequency range.  
The curve in Figure~2 is a fit to this form, 
with $\epsilon_{r} \equiv \epsilon / \epsilon_{0}$ = 1.72, 
$\omega_{0} = 300.6$~GHz, $\Gamma = 5.5$~GHz, and $F$ = 0.011.  The value 
of $F$ for oriented crystals would be much higher, since the electromagnetic 
pulse only stimulates transitions in crystals that are oriented with their 
$c$-axes parallel to the direction of propagation.  

Although we have used a standard Lorentzian lineshape with no
inhomogeneous broadening to fit these data, it is clear that an equal or
better fit could be achieved by considering inhomogeneous broadening. 
While we can confidently measure the linewidth, we cannot draw any
conclusions about the lineshape. However, there has been another
measurement of this absorption line by Mukhin {\it et al.}\cite{Mukhin}  
Their data strongly favor a Gaussian fit to the transmission with a full 
width at half maximum (FWHM) of 7~GHz.\cite{MukhinP}  (For this material, 
the FWHM of $ln(|\tilde{t}|)$, $Im(n)$, and $Im(\mu)$ are all approximately 
equal, so this value can be compared directly to $\Gamma$ in Eq. 2.)  The two 
measured widths are fairly close, but are not equal within the experimental 
uncertainty.  A possible cause of this difference will be discussed later 
in this paper.

This Gaussian fit is strong evidence for inhomogeneous broadening of the
linewidth.  However, the cause of this inhomogeneous broadening is
unclear. The most obvious source is local magnetic fields, which could
be due to the dipolar field of neighboring clusters or the nuclear
moments of the manganese atoms.  The local dipolar field has not, to our
knowledge, been calculated because it depends on the random orientations
along the $\pm z$ directions of the magnetic moments of all the other
clusters.  The calculated value of the hyperfine field varies depending
on the type of coupling that is assumed between the electron spins, but
its maximum possible value ranges from 270 to 
539~G,\cite{Hartmann-Boutron96} and its FWHM has been estimated at
280-380~G.\cite{Wernsdorfer99}

The effect of a local field would be to raise or lower the transition
energy between the $\pm 10$ and $\pm 9$ levels, $\Delta E_{10,9}$. From
the Hamiltonian, we see that $\Delta E_{10,9}(B)$ = $\Delta
E_{10,9}(B=0) \pm \mu_{B} g B_{z}$.  For this material, $g$ is very close
to 2, so to explain a linewidth of 5.5~GHz requires a magnetic field
distribution with width of 0.20~T.  

This local field distribution seems to be ruled out by the measurements
of the width of the tunneling peak performed by Friedman {\it et al.}
\cite{Friedman98}  Starting with a sample cooled in zero applied field,
a small magnetic field was applied parallel to the $z$-axis and the
relaxation rate of the magnetization toward its equilibrium value was
measured.  They measured the full width at half maximum to be 236~Oe at
2.6~K.  If the local field had a full width at half maximum of 0.20~T,
as implied above, then this narrow peak could never have been observed. 
In addition, the line shape measured by Friedman {\it et al.} was
clearly Lorentzian.  This implies that the inhomogeneous broadening due
to local magnetic fields must be significantly smaller than the
homogenous effect of lifetime broadening.  We note that Friedman 
{\it et al.}'s interpretation of this homogeneous broadening as being 
due to a lifetime of 250 ps does not imply that Lorentzian tails should 
be visible in the transitions between $m = \pm 10$ and $m = \pm 9$, since 
the lifetime of states with larger $|m|$ could be expected to be longer 
than that of the states involved in tunneling.  However, the absence 
of broadening of Friedman {\it et al.}'s data due to local magnetic 
fields is puzzling.

One possible explanation is that the width measured in tunneling
experiments reflects collective effects.  The relaxation rate measured
by Friedman {\it et al.} is a fit to the exponential relaxation that is
observed after an initial non-exponential relaxation.  As was pointed
out in Ref. \onlinecite{ProkofÕev}, relaxation must actually be a 
collective process in which the relaxation of one molecule changes the 
dipole fields of its neighbors, thereby bringing new molecules into 
resonance.  Therefore, it is possible that the width in the relaxation 
rate would be different from the instantaneous width measured in photon 
absorption experiments.

In fact, our calculations of the dipolar field show that collective
effects must decrease the width of this tunneling peak.  Using the
atomic positions tabulated in Ref. \onlinecite{Lis80}, we calculated the
effective field on one cluster due to each of its neighbors.  Because
the clusters have sizes that are not negligible compared to their
separations, this was done by calculating the dipolar field due to the
electronic spin of every Mn atom in the neighboring cluster with moment
${\bf m}_{j} = g \mu_{B} S_{j} \hat{\bf z}$, where we choose $S_j = +2$
for the eight Mn$^{3+}$ ions and $S_j = -3/2$ for the four Mn$^{4+}$ ions
to reflect their antiferromagnetic coupling. The fields due to all
twelve Mn ions located at positions ${\bf r}_{j}$ in the neighboring
cluster are summed at the position ${\bf r}_{i}$ of every Mn atom in
the cluster under consideration: 
\begin{equation}
{\bf B}_{i} = \sum_{j} \frac{\mu_0}{4 \pi} 
[\frac{3 ({\bf m}_{j} \cdot {\bf r}_{ij}) {\bf r}_{ij}}{r_{ij}^{5}} - 
\frac{{\bf m}_{j}}{r_{ij}^{3}}],
\label{Bi}
\end{equation}
where ${\bf r}_{ij}$ is the relative position ${\bf r}_{j} - {\bf r}_{i}$.
(We assume that the electronic spins are localized on the atomic position, 
ignoring the actual electronic density.)  An effective longitudinal field 
is then calculated: 
\begin{equation}
B_{\rm eff} = \frac{1}{m_{\rm total}} \sum_{i} {\bf B}_{i} 
\cdot {\bf m}_{i},
\label{Beff}
\end{equation}
where ${\rm m_{total}} = 10 g \mu_{B}$.

In order to correctly calculate the standard deviation of the dipolar field 
due to N neighbors exactly, it would be necessary to consider all $2^N$ 
possible arrangements of their spins.  However, we can obtain a lower limit 
on this standard deviation simply by considering the neighboring clusters that 
contribute the largest fields.  As shown in Table~1, there are only 
28 neighboring clusters that contribute an effective field 
greater than 10~G.  Assuming there is no ordering of the dipoles, these fields 
add randomly to give a FWHM of 520~G, as shown in Figure 3.  The combination 
of this field and the hyperfine field is not sufficient to explain the width 
observed in our spectroscopy. 
However, we note that it is considerably larger than the width of 236~Oe 
FWHM observed by Friedman {\it et al.}, implying that collective effects 
narrow the observed tunneling width.

\begin{table}
\caption{Fields due to neighboring clusters are calculated using 
Eqs.~~(\ref{Bi}) and ~(\ref{Beff}). All clusters that result 
in effective fields greater than 10 G are listed.  The relative 
location is given in terms of the unit cell dimensions.  \label{table1}}
\begin{ruledtabular}
\begin{tabular*}{\hsize}
{c@{\extracolsep{0ptplus1fil}}c@{\extracolsep{0ptplus1fil}}}
location&field (gauss)\\
\colrule
(0,0,$\pm$1)&95.9\\
($\pm$1,0,0)&56.9\\
(0,$\pm$1,0)&56.9\\
($\pm$0.5,$\pm$0.5,$\pm$0.5)&28.4\\
(0,0,$\pm$2)&19.2\\
($\pm$0.5,$\pm$0.5,$\pm$1.5)&18.6\\
($\pm$1,$\pm$1,0)&15.6\\
\end{tabular*}
\end{ruledtabular}
\end{table}

\begin{figure}
\includegraphics[width=\linewidth]{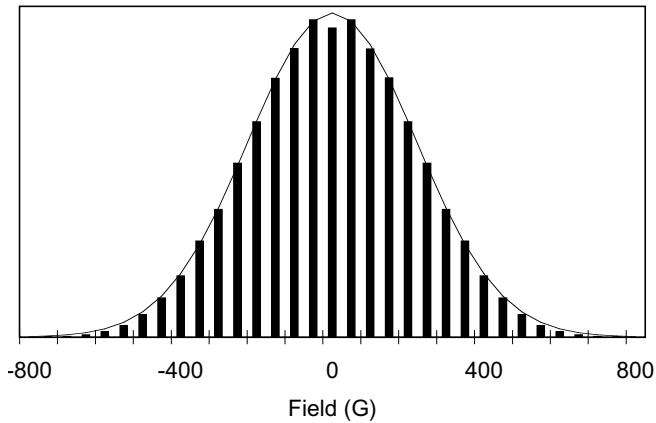}
\caption{Histogram of the distribution of fields resulting from all 
possible orientations of the neighboring clusters listed in Table~1.  
The line is a Gaussian with FWHM = 520 G.}
\end{figure}

One way we could understand the spectroscopy and the tunneling results
simultaneously is if there were some source of inhomogeneous broadening
other than local magnetic fields.  For example, if $\alpha$, the
anisotropy constant in the Hamiltonian, varied slightly ($\sim 1\%$)
between clusters due to small configurational changes of the ligands
surrounding them, then this variation would be seen in the linewidth of
the photon-induced transition, but not in the zero field tunneling.  The
photon absorption experiments would be sensitive to variations in
$\alpha$, but the width seen in the zero field tunneling experiments
would be due to homogeneous broadening mechanisms, either due to
the tunneling time itself or the interactions with phonons, as proposed
in Ref.~\onlinecite{Leuenberger00}. Calculations by Pederson {\it et
al.} suggest that the arrangement of the ligands is important in
determining the anisotropy energy.\cite{Pederson99} We note that
isomerism of the ligands has been shown to exist in these molecules and
results in large variations in the parameters in the Hamiltonian;
\cite{Aubin97,Aubin01,Wernsdorfer99} the variations in the ligand
positions responsible for the above effect would have to be less
dramatic.  It is possible that proximity to the isomers could cause
neighboring clusters to have some small variation in their Hamiltonians.  
We also note that recent work of Chudnovsky and Garanin postulates
dislocations as the source of spin tunneling and calculates the effect
of these dislocations on $\alpha$.\cite{Chudnovsky01}  The distribution 
in anisotropy constants calculated in that paper is in very good 
agreement with the distribution we observe in spectroscopic measurements.  
It is slightly narrower than the distribution observed by 
Mukhin {\it et al.}, but since the density of defects was chosen 
arbitrarily, this could simply imply that the defect density was higher 
in the sample measured by Mukhin {\it et al.}

Such a distribution of $\alpha$ would have no effect on the width of the 
relaxation peak in zero field, since in zero field all levels $\pm m$ are 
degenerate regardless of $\alpha$.  However, tunneling peaks at non-zero 
fields would be broadened, since (neglecting $\beta$) the field 
$H_{n} = -\alpha n / g \mu_{B}$ that brings the levels $m$ and 
$-m + n$ into resonance is proportional to $\alpha$.  It would be difficult 
to observe this broadening in tunneling measurements, since the term 
$\beta S_{z}^{4}$ brings different levels into resonance at different fields, 
so that if more than one level $m$ is involved in the thermally-assisted 
tunneling, the relaxation peak will be broadened.  However, we note that 
the width of the relaxation peaks in non-zero fields observed in Ref. 
\onlinecite{Zhong00} is sufficiently large to accommodate the required 
distribution of $\alpha$. 

In conclusion, we measure the linewidth of the single photon transition
from $m = \pm 10$ to $m = \pm 9$.  This linewidth, while surprisingly
large, is in agreement with that observed by Mukhin {\it et al.}
\cite{Mukhin}  The combination of this linewidth and that measured by
Friedman {\it et al.} \cite{Friedman98} suggests that the anisotropy 
constant varies between clusters.

\vspace{12pt}
We thank A. A. Mukhin for sharing his data and lineshape analysis.  
This research was supported by Colgate University and by an award from Research 
Corporation.  D.~N.~H. and G.~C. thank the NSF for support.


\begin{thebibliography}{Hartmann-Boutron96}

\bibitem{Friedman96} Jonathan R. Friedman, M.P. Sarachik, J. Tejada, and R.
Ziolo, Phys. Rev. Lett. {\bf 76}, 3830 (1996).

\bibitem{Thomas96} L. Thomas, F. Lionti, R. Ballou, D. Gatteschi, 
R. Sessoli, and B. Barbara, Nature {\bf 383}, 145 (1996).

\bibitem{Barra97} Anne Laure Barra, Dante Gatteschi, and Roberta Sessoli, 
Phys. Rev. B {\bf 56}, 8192 (1997).

\bibitem{Mirebeau99} I. Mirebeau, M. Hennion, H. Casalta, H. Andres, 
H. U. G\"{u}del, A. V. Irodova, and A. Caneschi, Phys. Rev. Lett. {\bf 83}, 
628 (1999).

\bibitem{Hill98} S. Hill, J. A. A. J. Perenboom, N. S. Dalal, T. Hathaway, 
T. Stalcup, and J. S. Brooks, Phys. Rev. Lett. {\bf 80}, 2453 (1998).

\bibitem{Fort98} A. Fort, A. Rettori, J. Villain, D. Gatteschi, and 
R. Sessoli, Phys. Rev. Lett. {\bf 80}, 612 (1998).

\bibitem{Luis98} Fernando Luis, Juan Bartolom\'{e}, and Julio F. 
Fern\'{a}ndez, Phys. Rev. B {\bf 57}, 505 (1998).

\bibitem{Bao01} W. Bao, R. A. Robinson, J. R. Friedman, H. Casalta, E. 
Rumberger, and D. N. Hendrickson, cond-mat/0008042.

\bibitem{Mukhin} A. A. Mukhin, V. D. Travkin, A. K. Zvezdin, A. Caneschi, 
D. Gatteschi, and R. Sessoli, Physica B, {\bf 284-288}, 1221 (2000) and 
A. A. Mukhin, V. D. Travkin, A. K. Zvezdin, S. P. Lebedev, A. Caneschi, 
and D. Gatteschi,  Europhys. Lett. {\bf 44}, 778 (1998).

\bibitem{Zhong99} Yicheng Zhong, M. P. Sarachik, Jonathan R. Friedman, 
R. A. Robinson, T. M. Kelley, H. Nakotte, A. C. Christianson, F. Trouw, 
S. M. J. Aubin, and D. N. Hendrickson, J. of Appl. Phys. 
{\bf 85}, 5636 (1999).

\bibitem{Lis80} T. Lis, Acta Crystallogr. B {\bf 36}, 2042 (1980).

\bibitem{Ibach} H. Ibach and H. L\"{u}th, {\it Solid-State Physics}, 
(Springer, Berlin, 1995), p. 291.

\bibitem{MukhinP} A. Mukhin, B. Gorshunov, M. Dressel, C. Sangregorio, 
and D. Gatteschi, Phys. Rev. B {\bf 63}, 214411 (2001), and 
A. Mukhin, private communication.

\bibitem{Hartmann-Boutron96} Fran\c{c}oise Hartmann-Boutron, Paolo Politi, 
and Jacques Villain, Int. J. of Mod. Phys. B {\bf 10}, 2577 (1996).

\bibitem{Wernsdorfer99} W. Wernsdorfer, R. Sessoli, and D. Gatteschi, 
Europhys. Lett. {\bf 47}, 254 (1999).

\bibitem{Friedman98} Jonathan R. Friedman, M. P. Sarachik, and R. Ziolo,
Phys. Rev. B {\bf 58}, R14729, (1998).

\bibitem{ProkofÕev} N. V. Prokofev and P. C. E. Stamp, Phys. Rev. Lett. 
{\bf 80}, 5794 (1998).

\bibitem{Leuenberger00} Michael N. Leuenberger and Daniel Loss, Phys. Rev. B 
{\bf 61}, 1286 (2000).

\bibitem{Pederson99} M. R. Pederson and S. N. Khanna, Phys. Rev. B {\bf 60}, 
9566 (1999).

\bibitem{Aubin97} S. M. J. Aubin, Ziming Sun, Ilia A. Guzei, Arnold L. 
Rheingold, George Christou, and David N. Hendrickson, Chem. Commun., 
2239 (1997).

\bibitem{Aubin01} Sheila M. J. Aubin, Ziming Sun, Hilary J. Eppley, Evan 
M. Rumberger, Ilia A. Guzei, Kirsten Folting, Peter K. Gantzel, Arnold L. 
Rheingold, George Christou, and David N. Hendrickson, Inorganic Chemistry, 
{\bf 40}, 2127 (2001).

\bibitem{Chudnovsky01} E. M. Chudnovsky and D. A. Garanin, cond-mat/0105195 
and cond-mat/0105518.

\bibitem{Zhong00} Yicheng Zhong, M.P. Sarachik, Jae Yoo, and D. N. Hendrickson,
Phys. Rev. B {\bf 62}, R9256 (2000).


\end{thebibliography}
\end{document}